# Terracentric Nuclear Fission Georeactor: Background, Basis, Feasibility, Structure, Evidence, and Geophysical Implications

J. Marvin Herndon
Transdyne Corporation
San Diego, CA 92131 USA
mherndon@san.rr.com

**Abstract:** The background, basis, feasibility, structure, evidence, and geophysical implications of a naturally occurring Terracentric nuclear fission georeactor are reviewed. For a nuclear fission reactor to exist at the center of the Earth, all of the following conditions must be met: (1) There must originally have been a substantial quantity of uranium within Earth's core; (2) There must be a natural mechanism for concentrating the uranium; (3) The isotopic composition of the uranium at the onset of fission must be appropriate to sustain a nuclear fission chain reaction; (4) The reactor must be able to breed a sufficient quantity of fissile nuclides to permit operation over the lifetime of Earth to the present; (5) There must be a natural mechanism for the removal of fission products; (6) There must be a natural mechanism for removing heat from the reactor; (7) There must be a natural mechanism to regulate reactor power level; and; (8) The location of the reactor must be such as to provide containment and prevent meltdown. Herndon's georeactor alone is shown to meet those conditions. Georeactor existence evidence based upon helium measurements and upon antineutrino measurements is described. Geophysical implications discussed include georeactor origin of the geomagnetic field, geomagnetic reversals from intense solar outbursts and severe Earth trauma, as well as georeactor heat contributions to global dynamics. The article is organized as follows: 1.0 Introduction and Background; 2.0 Georeactor Basis; 2.1 Uranium in Earth's Core; 2.2 Uranium Concentration; 2.3 Georeactor Nuclear Calculations; 2.4 Requisite Uranium Isotopic Composition; 2.5 Requisite Fuel Breeding; 2.6 Requisite Fission Product Removal; 2.7 Requisite Georeactor Heat Removal; 2.8 Requisite Regulation Mechanism; 2.9 Requisite Georeactor Containment; 3.0 Georeactor Existence Evidence Based on Helium Measurements; 4.0 Georeactor Existence Evidence Based on Antineutrino Measurements; 5.0 Heat Flow Considerations and Georeactor Contributions to Geodynamics; 6.0 Georeactor Origin of the Geomagnetic Field; 7.0 Georeactor Geomagnetic Reversals; 8.0 Perspectives, followed by Acknowledgements and References.

## 1.0 Introduction and Background

The 1938 discovery of nuclear fission [1, 2], the splitting of the uranium nucleus, fundamentally changed human perceptions on warfare, energy production, and the nature of planet Earth. Much has been written on the subject of nuclear fission weapons and power plants; comparatively little has been written about terrestrial natural nuclear fission. The latter is the subject of this review.

In 1939, Libby attempted to discover whether uranium in nature undergoes spontaneous nuclear fission [3]. His negative result implied that, if uranium could decay by spontaneous fission, the spontaneous fission half-life would be greater than $10^{14}$ years. In 1940, Flerov and Petrzhak [4, 5] announced the discovery of spontaneous nuclear fission in uranium with a half-life of about $10^{17}$ years.

During this period and extending to 1950, there was discussion of the possibility of large-scale nuclear reactions in the Earth's crust or mantle [6-8]. In 1953, Fleming and Thode [9] and Wetherill [10] studied the isotopic compositions of krypton and xenon extracted from uranium minerals and discovered that the fissionogenic isotope assemblages could be understood as a binary mixture of components from spontaneous fission and neutron-induced fission. Notably, in samples containing prodigious neutron absorbers, the neutron-induced component was low or absent, whereas in older samples with less neutron absorbers, the relative amount of neutron-induced fission in the mixture was significantly greater. In fact, in one sample of Belgian Congo pitchblend, Wetherill and Ingram [11] reported a 35% neutron-induced fission proportion in that mixture, leading them to state: "Thus the deposit was twenty-five percent of the way to becoming a pile [nuclear fission reactor]. It is also interesting to extrapolate back 2,000 million years where the $^{235}$U abundance was 6 percent instead of 0.7. Certainly, such a deposit would be closer to being an operating pile".

In 1956, Kuroda [12, 13] applied Fermi's nuclear reactor theory [14] and demonstrated the feasibility that 2,000 million years ago seams of uranium ore one meter in thickness could engage in neutron-induced nuclear fission chain reactions. He predicted that ground water would serve as a moderator slowing neutrons to thermal energy levels. For sixteen years, Kuroda later told me, the subject of natural nuclear fission reactors was unpopular in the geoscience community. In fact, he related, the only way those papers got published at all was that at the time the *Journal of Chemical Physics* would publish short papers without peer-review. But then, reality struck: Kuroda's prediction was proven to have taken place in nature.

In 1972, scientists at the French Commissariat à l' Énergie Atomique announced the discovery of the intact remains of a natural nuclear fission reactor in a uranium mine (Figure 1) at Oklo, near Franceville in the Republic of Gabon in Western Africa. Seams of uranium ore, one meter in thickness, had undergone sustained nuclear fission chain reactions 1,800 million years ago. The reactor had operated at a low power level, 10-100 kilowatts for a period of several thousand years. Control of the chain reaction appears to have been achieved by the reactor boiling off of ground water, the moderator for slowing neutrons, effectively shutting down the reaction, which restarted when the cooler environment allowed water to return. Although the chain reaction primarily involved slow (thermal) neutron fission, examination of the fission products showed the reactor had functioned to a lesser extend as a fast neutron breeder reactor, producing additional fissile elements from $^{238}$U.

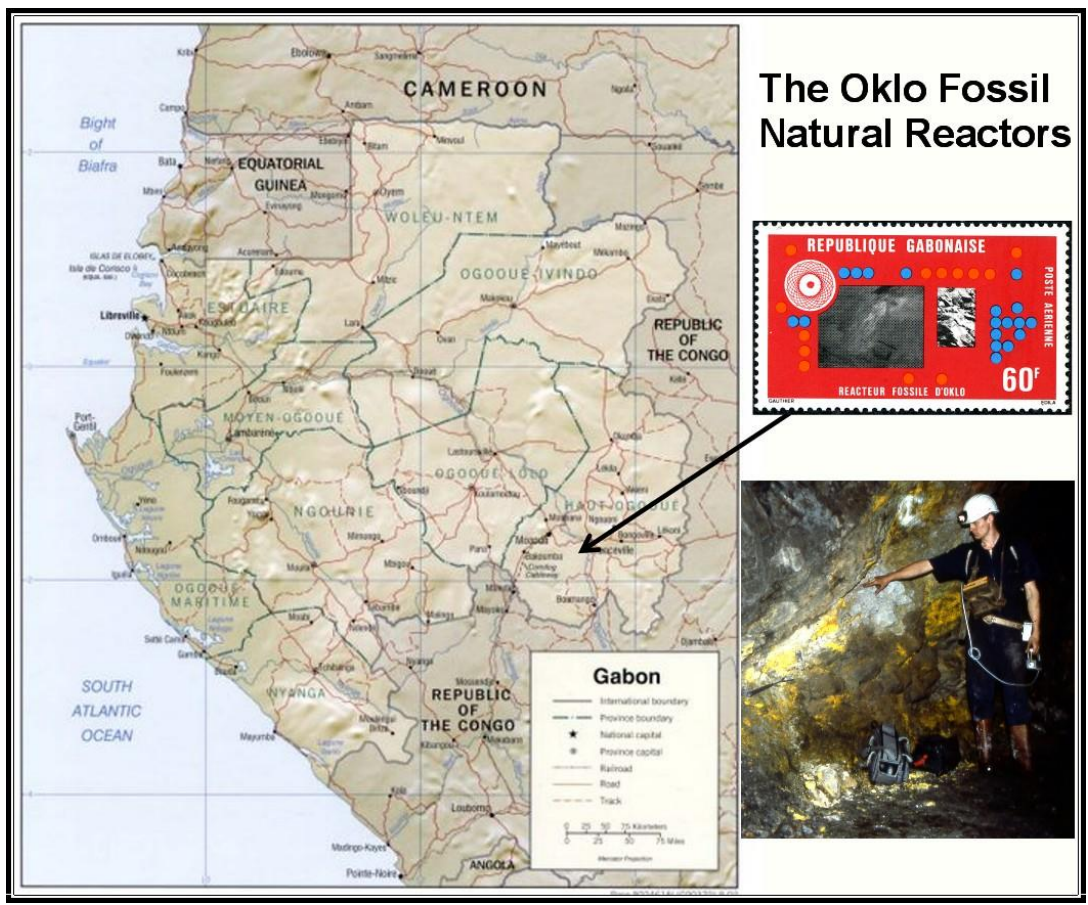

**Figure 1**. Location of the natural reactors at Oklo, near Franceville, in the Republic of Gabon in Western Africa, indicated by the commemorative postage stamp issued by that nation in honor of those natural reactors. Inset photo of a uranium seam in the reactor zone courtesy of Francoise Gauthier-Lafaye.

The discovery of fossil nuclear reactors at Oklo represented a profound revelation in human thought: Nuclear fission reactors are very much a part of nature, not simply man-made contrivances [15-19]. Significantly, Oklo investigations helped to pave the way for the next important advance, my demonstration of the feasibility of a nuclear fission reactor at the center of Earth.

The Earth has near or at its center a powerful energy source that powers the mechanism that generates the geomagnetic field. In 1993, I published a scientific paper entitled "Feasibility of a Nuclear Fission Reactor at the Center of the Earth as the Energy Source for the Geomagnetic Field" [20]. In that paper I used an approach similar to that employed by Kuroda [12, 13], demonstrating the feasibility based upon application of Fermi's nuclear reactor theory [14]. But unlike Kuroda, who had knowledge of uranium deposits on Earth's surface, I had to provide justification for significant uranium occurring in the Earth's core and to provide a mechanism in nature for its concentration. In the intervening years, I developed the concept, understanding better the nature, structure, and geophysical consequences which I described in a series of scientific papers [20-28] and books [29-31]. Here I review the subject of Earth's nuclear fission georeactor, which is shown schematically in Figure 2.

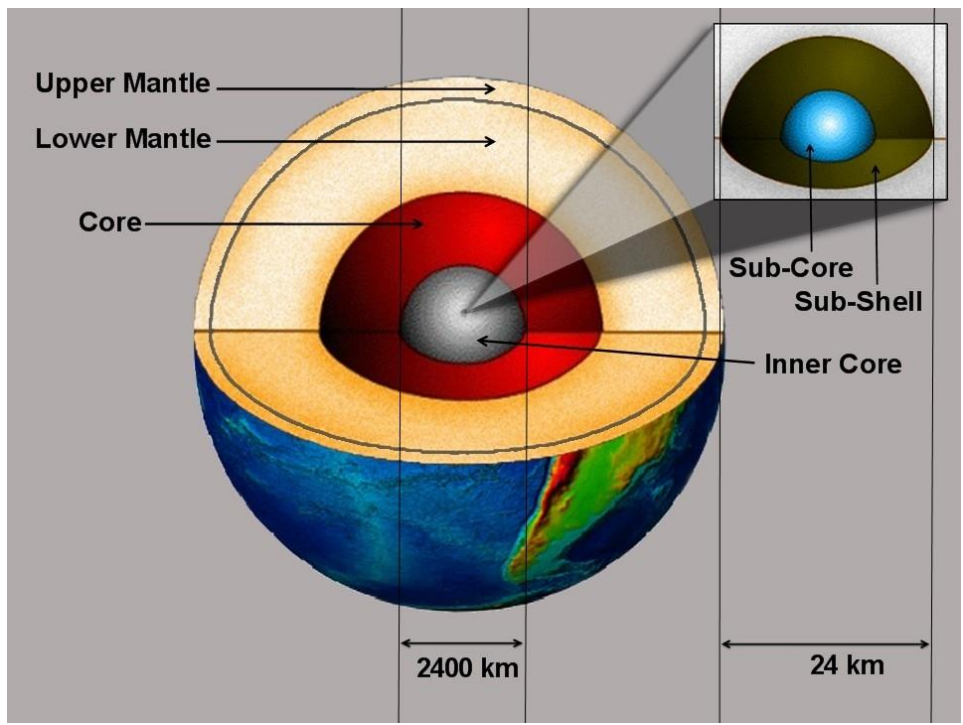

**Figure 2.** Earth's nuclear fission georeactor (inset) shown in relation to the major parts of Earth. The georeactor at the center is one ten-millionth the mass of Earth's fluid core. The georeactor sub-shell, I posit, is a liquid or a slurry and is situated between the nuclear-fission heat source and inner-core heat sink, assuring stable convection, necessary for sustained geomagnetic field production by convection-driven dynamo action in the georeactor sub-shell [22, 26, 27].

Fundamental concepts from the 1930s and 1940s underpin the current textbook explanation of Earth's structure and composition. Briefly, these are: (1) The Earth resembles an ordinary chondrite meteorite; (2) the inner core is iron metal in the process of solidifying from the fluid iron alloy core; and, (3) the silicate mantle is of uniform chemical composition with its observed seismic discontinuities explained as boundaries between pressure-induced changes in crystal structure. Forty years after the inner core's composition was pronounced, I published a different idea for its composition which led me to show, by fundamental ratios of mass, that Earth resembles, not an ordinary chondrite, but a highly reduced enstatite chondrite which implied different deep-Earth chemistry. One chemical consequence is that copious amounts of uranium exist within the core, instead of exclusively in the mantle as previously thought. Subsequently, I demonstrated the feasibility of a Terracentric nuclear fission reactor, and then developed the concept, which is connected to the production of the geomagnetic field, and is a major new energy source with significant geological implications, for example, related to hotspots such as underlies the Hawaiian Islands. I did these things singlehandedly, so this review of Terracentric nuclear fission is quite properly a review of my own work in the subject area.

## 2.0 Georeactor Basis

For a nuclear fission reactor to exist at the center of the Earth, all of the following conditions must be met:

- There must originally have been a substantial quantity of uranium within Earth's core.
- There must be a natural mechanism for concentrating the uranium.
- The isotopic composition of the uranium at the onset of fission must be appropriate to sustain a nuclear fission chain reaction.
- The reactor must be able to breed a sufficient quantity of fissile nuclides to permit operation over the lifetime of Earth to the present.
- There must be a natural mechanism for the removal of fission products.
- There must be a natural mechanism for removing heat from the reactor.
- There must be a natural mechanism to regulate reactor power level.
- The location of the reactor must be such as to provide containment and prevent meltdown.

In the following subsections, I describe the manner by which each of the above conditions is fulfilled for the Herndon's nuclear fission georeactor at the center of Earth, and not fulfilled for other, later, putative 'georeactors' assumed to be located elsewhere in Earth's deep interior.

## 2.1 Uranium in Earth's Core

In 1898, Wiechert suggested that the Earth's mean density could be accounted for if Earth has a core made of nickeliferous iron metal, like the iron meteorites he had seen in museums [32]. In 1906, Oldham discovered that earthquake-wave velocities increase with depth, but then slow abruptly; he had discovered the Earth's core [33]. In 1936, Lehmann discovered the inner core by reasoning that its existence could account for earthquake waves being reflected into a region where they should otherwise not have been detected [34]. But what is the chemical composition of the inner core?

Studies of earthquake waves and moment of inertia calculations can delineate structures within the earth and their physical states, but not their chemical compositions. For compositions, one must rely upon implications derived from chondrite meteorites. Chondrite elements, like corresponding elements in the outer portion of the Sun, were never appreciably separated from one another; they thus provide a basis for understanding the bulk composition of Earth. But the situation is complicated because there are three groups of chondrites (*ordinary*, *carbonaceous*, and *enstatite* chondrites) that differ significantly in their states of oxidation and in their mineral assemblages [35].

Of the three groups of chondrites, only ordinary chondrites and enstatite chondrites contain appreciable quantities of iron metal. But enstatite chondrites are quite rare and the origin of their highly reduced state of oxidation was not understood. So, circa 1940, there was widespread perception that the Earth resembles an ordinary chondrite. The inner core was thought to be partially crystallized nickel-iron metal in the process of solidifying by freezing from the Earth's nickel-iron core [36]. That conclusion was reached because in ordinary chondrites nickel is invariably alloyed with iron metal, and elements heavier than nickel are insufficiently abundant to account for a mass as great as the inner core.

Subsequent seismic investigations revealed boundaries just above the core and also in the mantle, 660 km below the Earth's surface, where earthquake waves impinging at an angle change speed and direction. Several such seismic boundaries were also discovered in the upper mantle. Seismic boundaries or discontinuities can potentially arise from two very different causes. Earthquake waves change speed and direction when passing from one substance into a chemically different substance or when traversing the boundary between two different crystal structures of the same material. The former made no sense under the assumption that the Earth is similar to an ordinary chondrite, so pressure-induced crystalline phase boundaries became the explanation that is widely believed even now. But the Earth resembles, not an ordinary chondrite, but an enstatite chondrite and the seismic discontinuities at the depth of 660 km and below characterize the boundaries between different chemical compositions of matter.

Imagine heating a metal-bearing chondrite. At some temperature below the melting points of the silicate minerals, the iron sulfide melds with the nickel-iron metal and forms a liquid capable of percolating downward by gravity forming a two component system analogous to the structure of Earth. Figure 3 shows that only enstatite chondrites, not ordinary chondrites, harbor a sufficient proportion of iron alloy to account for the massive core of Earth.

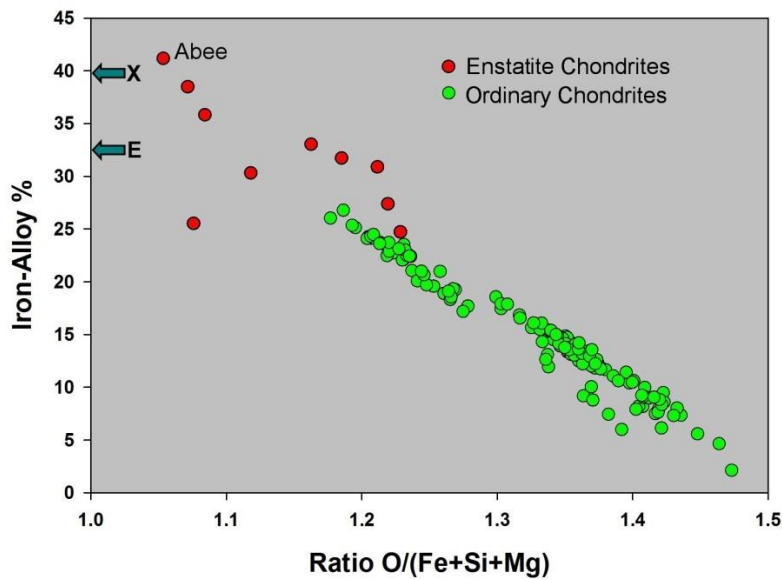

**Figure 3.** Evidence that Earth resembles an enstatite chondrite. The percent alloy (iron metal plus iron sulfide) of 157 ordinary chondrites (green circles) and 9 enstatite chondrites (red circles) plotted against oxygen content. The core percent of the whole-Earth, "arrow E", and of (core-plus-lower mantle), "arrow X", shows that Earth resembles an Abee-type enstatite chondrite and does not resemble an ordinary chondrite. Data from references [37-40].

Forty years after Birch explained the inner core's composition as being partially crystallized nickel-iron metal, I deduced its composition as fully crystallized nickel silicide [41] based upon discoveries made in the 1960s [42-44]. Subsequently, I discovered that the mass ratios of the components of the inner 82% of Earth are virtually identical to corresponding components of a primitive enstatite chondrite, as shown in Table 1. That identity means that the components of a primitive enstatite chondrite are compositionally similar to corresponding components in Earth's deep interior. Moreover, it means the deep interior of the Earth has the same highly reduced state of oxidation as a primitive enstatite chondrite. Furthermore, it means that a substantial quantity of uranium occurs in the Earth's core as most, if not all, of the uranium in the primitive Abee enstatite chondrite occurs in the part that corresponds to the Earth's core [45].

**Table 1.** Fundamental mass ratio comparison between the endo-Earth (lower mantle plus core) and the Abee enstatite chondrite. Above a depth of 660 km, seismic data indicate layers suggestive of veneer, possibly formed by the late addition of more oxidized chondrite and cometary matter, whose compositions cannot be specified with certainty at this time.

| Fundamental Earth Ratio | Earth Ratio Value | Abee Ratio Value |
|---|---|---|
| lower mantle mass to total core mass | 1.49 | 1.43 |
| inner core mass to total core mass | 0.052 | theoretical 0.052 if $Ni_3Si$ 0.057 if $Ni_2Si$ |
| inner core mass to lower mantle + total core mass | 0.021 | 0.021 |
| D″ mass to total core mass | 0.09*** | 0.11* |
| ULVZ** of D″ CaS mass to total core mass | 0.012**** | 0.012* |

\* = avg. of Abee, Indarch, and Adhi-Kot enstatite chondrites
D″ is the "seismically rough" region between the fluid core and lower mantle
\*\* ULVZ is the "Ultra Low Velocity Zone" of D″
\*\*\* calculated assuming average thickness of 200 km
\*\*\*\* calculated assuming average thickness of 28 km
data from [46-48]

## 2.2 Uranium Concentration

Only five elements comprise about 95% of the mass of a chondrite meteorite and by inference the Earth; the four minor elements add about 3%. Figure 4 shows the distribution of those elements between the alloy and the silicate portions of a primitive enstatite chondrite, and, by the identity shown in Table 1, between the core and mantle of Earth. As consequence of the high state of reduction, certain elements that have a high affinity for oxygen, including calcium, magnesium and silicon, occur in part in the iron alloy portion. The high state of reduction, a consequence of Earth's formation [25, 27, 49, 50], is the reason that uranium occurs in the Earth's core.

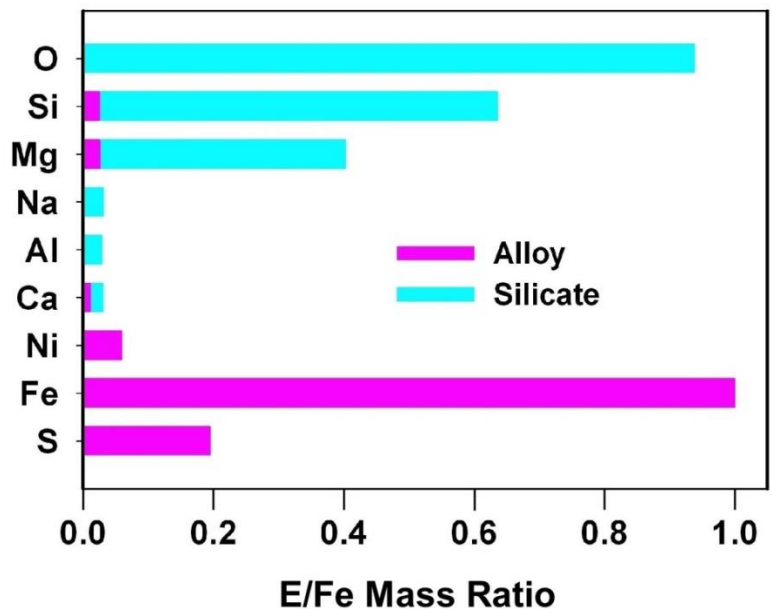

**Figure 4.** Relative abundances of the major and minor elements in the Abee enstatite chondrite, normalized to iron, showing their relative amounts in the alloy and silicate portions. Note that calcium (Ca), magnesium (Mg), and silicon (Si), normally lithophile elements, occur in part in the alloy portion.

Elements that have a high affinity for oxygen are generally incompatible in an iron alloy. Upon cooling from a high temperature these oxyphile elements escape the iron alloy by precipitating when thermodynamically possible. In the Earth's core, calcium and magnesium reacted with sulfur at a high temperature to form CaS and MgS, which floated to the top of the core. Silicon precipitated by combining with nickel. The nickel silicide sank by the action of gravity and formed the inner core.

In the deep interior of Earth, density is a function almost exclusively of atomic number and atomic mass. The core is layered on the basis of density. Uranium, more dense than the inner core by more than a factor of two, either as metal or mono-sulfide, driven by gravity, perhaps through a series of steps, concentrated at the gravitational center of Earth. Figure 5, a schematic representation of the deep interior of Earth, illustrates the layering by density which, with the data from Table 1, explains well the observed seismic boundaries as compositional boundaries, not pressure-induced crystalline phase boundaries.

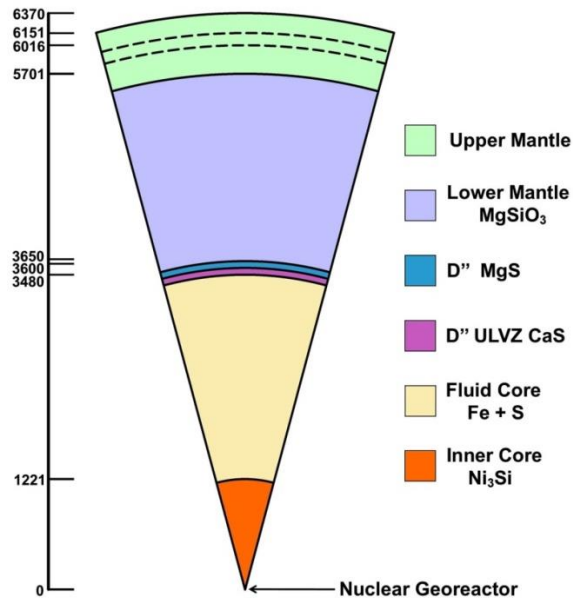

**Figure 5.** Schematic representation of the layers of the Earth based upon the data from which Table 1 is derived. Scale in km.

### 2.3 Georeactor Nuclear Calculations

In his nuclear reactor theory, Fermi [14] defined the condition for a self-sustaining nuclear fission chain reaction. The value of $k_{eff}$ ($k$ effective) represents the number of fission neutrons in the present population divided by the number of fission neutrons in the previous population. The defining condition for self-sustaining nuclear fission chain reactions is that $k_{eff} = 1.0$. If $k_{eff} > 1.0$, the neutron population and the energy output are increasing and will continue until changes in the fuel, moderators, and neutron absorbers cause $k_{eff}$ to decrease to 1.0. If $k_{eff} < 1.0$, the neutron population and energy output are decreasing and will eventually decrease to zero. If $k_{eff}$ = 1.0, the neutron population and energy output are constant.

Initially, I demonstrated the feasibility of a Terracentric nuclear fission georeactor by applying Fermi's nuclear reactor theory. Subsequently, far more sophisticated numerical simulation calculations were made using the SAS2 analysis sequence contained in the SCALE Code Package from Oak Ridge National Laboratory [51] that has been developed over a period of three decades and has been extensively validated against isotopic analyses of commercial reactor fuels [52-56]. The SAS2 sequence invokes the ORIGEN-S isotopic generation and depletion code to calculate concentrations of actinides, fission products, and activation products simultaneously generated through fission, neutron absorption, and radioactive decay. The SAS2 sequence performs the 1-D transport analyses at selected time intervals, calculating an energy flux spectrum, updating the time-dependent weighted cross-sections for the depletion analysis, and calculating the neutron multiplication of the system.

The difference between calculations based upon Fermi's nuclear reactor theory and the numerical simulation codes developed at Oak Ridge National Laboratory is similar to the difference between a photograph and a full-length motion picture. For a given configuration of fissionable elements, Fermi's theory allows only determination of the feasibility of a nuclear fission chain reaction at a single moment in time. The Oak Ridge simulations, on the other hand, progress through time in a series of discrete steps. At each step the nuclear reaction induced changes in fuel composition, which affect the next step, are calculated. Nuclear reaction induced changes include neutron-induced nuclear reactions, production of fission products, and the natural radioactive decay of fuel and fission products.

The Oak Ridge simulations were a major advance for georeactor calculations because they demonstrated that the georeactor could function over the entire age of the Earth through natural fuel-breeding nuclear reactions. The simulations yield quantitative estimates of specific fission products, one being helium, which serves as a geochemical tracer and offers an explanation for exhaled deep-Earth helium.

## 2.4 Requisite Uranium Isotopic Composition

Natural uranium consists mainly of the readily-fissionable $^{235}$U and the essentially non-fissionable $^{238}$U. In a natural nuclear fission reactor, the value of $k_{eff}$ is strongly dependent upon the ratio $^{235}$U/$^{238}$U. As shown by curves A, B, and C in Figure 6, a substantial mass of natural uranium (*e.g.*, a few hundred kilograms) during the first 2 gigayears of Earth's existence would be capable of undergoing sustained nuclear fission chain reactions. After that point in time, other conditions must also be fulfilled, namely, fuel breeding and fission product removal.

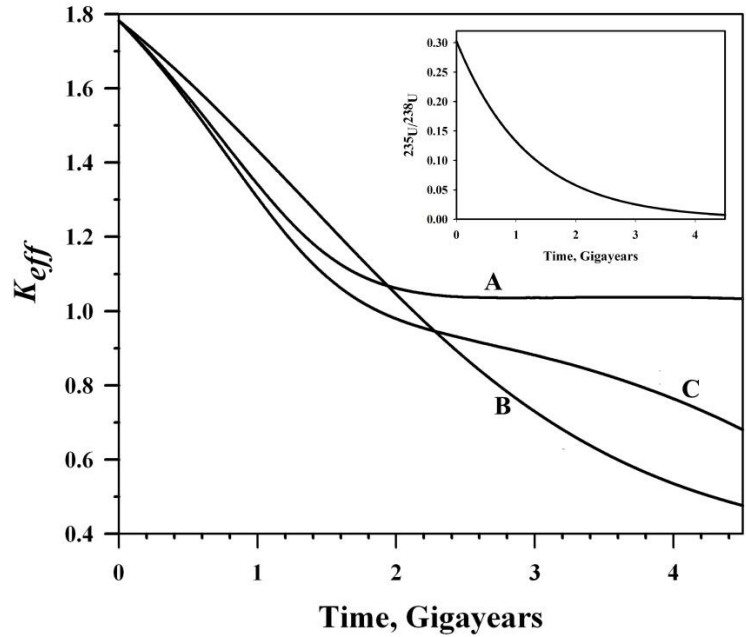

**Figure 6.** Numerical simulation results, chosen to illustrate main georeactor operational parameters and uncertainties, are presented in terms of $k_{eff}$ over the lifetime of the Earth. The curve labeled A is a 3 TW run in which fuel breeding occurs for the case of fission products instantly removed upon formation. Curve B is the same as A, except fission products are left in place. Curve C is a very low-power run, fission products instantly removed, where the low-level of breeding is insufficient for maintaining criticality. These show the importance of breeding, fission-product removal, and intrinsic self-regulation. Inset shows the natural decay of non-fissioning uranium over the lifetime of Earth.

## 2.5 Requisite Fuel Breeding

The half-life of $^{235}U$ is shorter than that of $^{238}U$. So, radioactive decay causes the ratio $^{235}U/^{238}U$ to decrease over time, as shown by the inset in Figure 6. Curves B and C in the same figure show that at some point during the decline, neutron absorption by the increasingly greater relative proportion of $^{238}U$ causes $k_{eff} < 1.0$, essentially killing the nuclear fission chain reaction, but curve A maintains criticality, *i.e.*, $k_{eff} > 1.0$, by fuel breeding reactions throughout the lifetime of Earth.

Under appropriate operating conditions, the neutrons produced by nuclear fission can "breed" additional fuel from essentially non-fissionable $^{238}$U. As noted by Seifritz [57] and I [20], the principal georeactor fuel-breeding takes place by the reaction

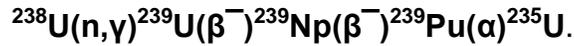

$$^{238}U(n,\gamma)^{239}U(\beta^-)^{239}Np(\beta^-)^{239}Pu(\alpha)^{235}U.$$

Curve B in Figure 6, calculated at a very low power level, shows the decrease in $k_{eff}$ that would result with too little fuel breeding. Curves A and B were each calculated with fission products promptly removed from the reactor zone.

## 2.6 Requisite Fission Product Removal

During the first two gigayears of Earth's existence, a deep-Earth nuclear fission georeactor could function without fuel breeding and without removal of fission products. But to maintain the nuclear fission chain reaction into the present requires both fuel breeding and prompt removal of fission products. Even with fuel breeding, $k_{eff}$ will diminish unless fission products are promptly removed. The calculations for curves A and C of Figure 6 were identical, except that fission products were left in place for curve C.

There is a natural process for the removal of fission fragments from a nuclear fission georeactor operating at the center of the Earth. In the deep interior of the Earth, density is a function almost entirely of atomic number and atomic mass. When the uranium nucleus fissions, it usually splits into two roughly equal parts, each part having approximately half the atomic number and half the atomic mass of the parent uranium. That means that the density of the fission fragments is less than the density of uranium. Consequently, the fission fragments tend to migrate outward and away from the reactor zone, while the uranium tends to re-concentrate downward.

## 2.7 Requisite Georeactor Heat Removal

The Terracentric georeactor produces heat through nuclear fission and through the decay of radionuclides. There is a natural process for heat removal from the georeactor. The georeactor is thought to have a two-part structure as illustrated schematically in Figure 7. The concentration of uranium at the center of the Earth forms the nuclear fission reactor zone, called the georeactor sub-core; this is where the heat is primarily generated by nuclear fission and by the natural decay of actinide elements.

Surrounding the sub-core is a shell composed of fission products and the products of radioactive decay, which is referred to as the georeactor sub-shell. Heat is also produced in the georeactor sub-shell by fission-product radioactive decay, although less heat than in the sub-core. The sub-shell is thought to be a liquid or slurry that is engaged in thermal convection [22, 26, 27].

The natural configuration of the georeactor is ideal for heat transport by thermal convection. Heat is produced primarily in the georeactor's nuclear sub-core, which heats the matter at the base of the georeactor's nuclear waste sub-shell causing it to expand, becoming less dense. The less dense 'parcel' of bottom matter floats to the top of the sub-shell where it contacts the massive inner core heat-sink and loses its extra heat, densifies, and sinks. The inner core heat-sink is surrounded by an even more massive heat-sink, the fluid iron alloy core, which helps to ensure the existence of an adverse temperature gradient in the georeactor sub-shell, a necessary condition for thermal convection.

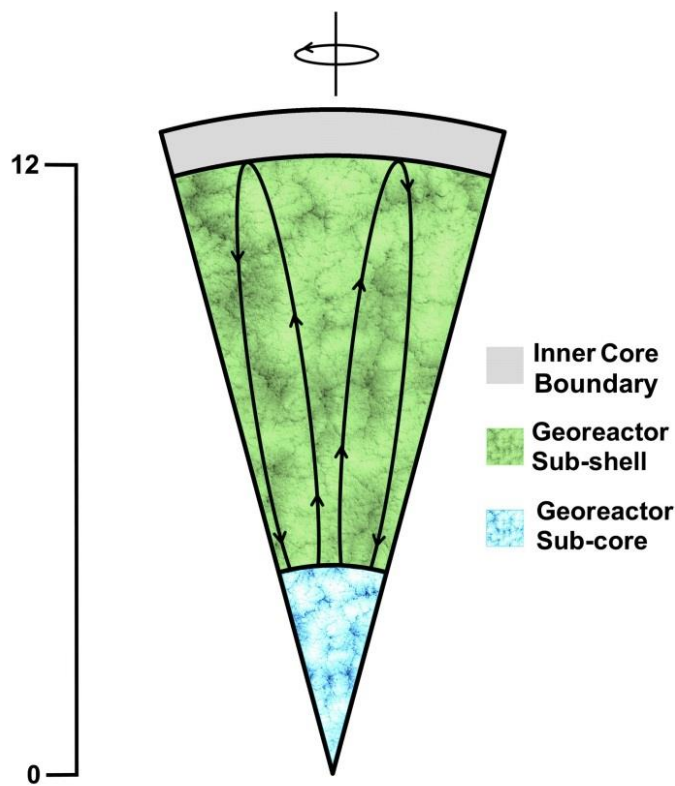

**Figure 7.** Schematic representation of the georeactor. Planetary rotation and fluid motions are indicated separately; their resultant motion is not shown. Stable convection with adverse temperature gradient and heat removal is expected. Scale in km.

## 2.8 Requisite Regulation Mechanism

During the first two gigayears of Earth's existence, as indicated in Figure 6, the $^{235}U/^{238}U$ ratio was quite large; initially the readily fissionable $^{235}U$ comprised 25% of the uranium. So, a highly energetic georeactor was possible in principle, unless the natural configuration of the georeactor affords a self-regulation mechanism.

Georeactor numerical simulations undertaken by Oak Ridge National Laboratory and by Transdyne Corporation were made at constant power levels. Assuming that the georeactor consisted of the maximum amount of available uranium [45], we determined a maximum constant power level of 30 terawatts (TW) [58]. At a higher power level, the georeactor would have fully consumed its fuel and ceased to operate before now. Similarly, too low a power level would be insufficient for the fuel breeding that is necessary for sustained georeactor operation during the last two gigayears (Figure 6, curve B). There must exist a natural georeactor self-regulation mechanism.

Reference Figure 7: In the micro-gravity environment at the center of Earth, georeactor heat production that is too energetic would be expected to cause actinide sub-core disassembly, mixing actinide elements with neutron-absorbers of the nuclear waste sub-shell, quenching the nuclear fission chain reaction. But as actinide elements begin to settle out of the mix, the chain reaction would restart, ultimately establishing a balance, a dynamic equilibrium between heat production and actinide settling-out, a self-regulation control mechanism [27].

## 2.9 Requisite Georeactor Containment

The georeactor's dense nuclear sub-core, described above, requires containment which is provided naturally by its location at Earth's center. Heat produced by nuclear fission chain reactions cannot cause meltdown as the sub-core already resides at the gravitational bottom. In the wake of interest stimulated by Herndon's georeactor, several attempts were made to describe 'georeactors' at locations other than at Earth's center, namely, at the core mantle boundary [59] and atop the inner core [60, 61]. But in each case there is no confinement. If nuclear fission chain reactions were to have occurred in those places, meltdown would inevitably take place and those putative 'georeactors' would meltdown to Earth's center, the location of Herndon's georeactor.

## 3.0 Georeactor Existence Evidence Based on Helium Measurements

When a uranium nucleus undergoes nuclear fission, it usually splits into two roughly equal, large fragments. Once in every 10,000 fission events, however, the nucleus splits into three pieces, two large and one very small. Tritium, $^3$H, is a prominent very small fragment of ternary fission. Tritium is radioactive with a half-life of 12.32 years and decays to $^3$He; $^4$He is likewise georeactor produced and also derives from the alpha particles of natural decay. Figure 8 presents helium fission product results from georeactor numerical simulations conducted at Oak Ridge National Laboratory, expressed as $^3$He/$^4$He relative to the same ratio measured in air [24]. That georeactor-produced $^3$He/$^4$He ratios have the same range of values observed in oceanic lava is strong evidence that the georeactor exists and is the source of the observed deep-Earth helium [62].

In 1969, Clarke et al. [63] discovered that $^3$He and $^4$He are venting from the Earth's interior. At the time there was no known deep-Earth mechanism that could account for the experimentally measured $^3$He, so its *ad hoc* origin was assumed to be a primordial $^3$He component, trapped at the time of Earth's formation, which was subsequently diluted with the appropriate amount of $^4$He from radioactive decay.

The $^3$He/$^4$He ratio of helium occluded in basalt at mid-ocean ridges is 8.6 ± 1 times greater than the same ratio in air, expressed as 8.6 $R_A$. Table 2 shows the commonality of normalized $^3$He/$^4$He ratios in the 2σ confidence interval for helium trapped in basalt extruded from undersea volcanoes.

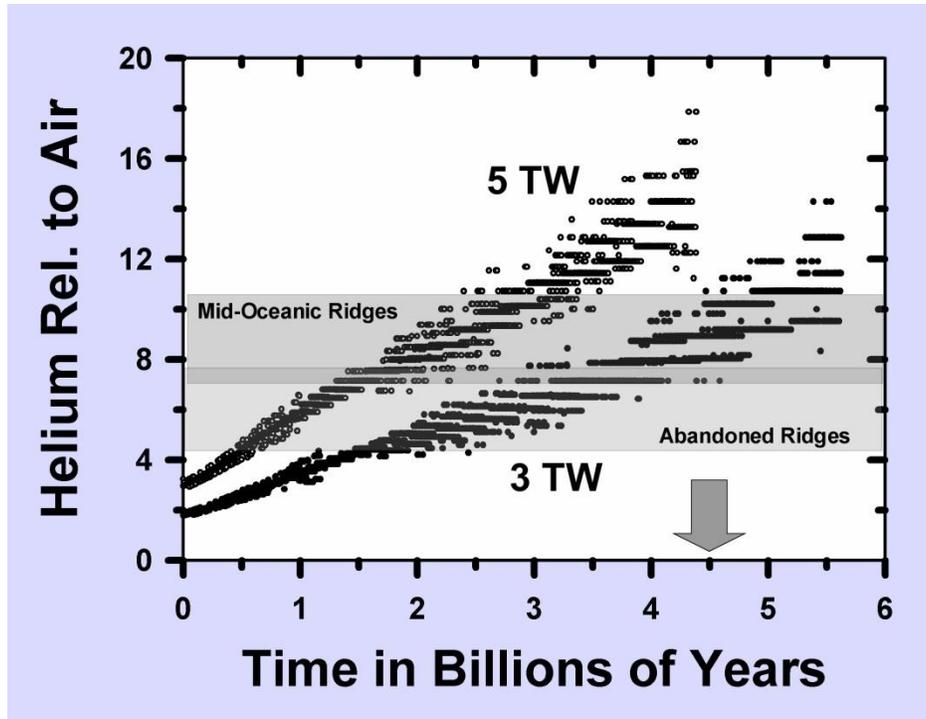

**Figure 8.** Fission product ratio $^3$He/$^4$He, relative to that of air, $R_A$, from nuclear georeactor numerical calculations at 5 terawatts, TW, (upper) and 3 TW (lower) power levels [24]. The band for measured values from mid-oceanic ridge basalts is indicated by the solid lines. The age of the Earth is marked by the arrow. Note the distribution of calculated values at 4.5 billion years, the approximate age of the Earth. The increasing values are the consequence of uranium fuel burn-up. Icelandic deep-Earth basalts present values ranging as high as 37 times the atmospheric value [64].

**Table 2**. Statistics of $^3$He/$^4$He relative to air ($R_A$) of basalts from along the global spreading ridge system at a two standard deviation ($2\sigma$) confidence level. Adapted from reference [65].

| Submarine Basalt Province | $^3$He/$^4$He Relative to Air ($R_A$) |
|---|---|
| Propagating Lithospheric Tears | 11.75 ± 5.13 $R_A$ |
| Manus Basin | 10.67 ± 3.36 $R_A$ |
| New Rifts | 10.01 ± 4.67 $R_A$ |
| Continental Rifts or Narrow Oceans | 9.93 ± 5.18 $R_A$ |
| South Atlantic Seamounts | 9.77 ± 1.40 $R_A$ |
| MORB | 8.58 ± 1.81 $R_A$ |
| EM Islands | 7.89 ± 3.63 $R_A$ |
| North Chile Rise | 7.78 ± 0.24 $R_A$ |
| Ridge Abandoned Islands | 7.10 ± 2.44 $R_A$ |
| South Chile Rise | 6.88 ± 1.72 $R_A$ |
| Central Atlantic Islands | 6.65 ± 1.28 $R_A$ |
| HIMU Islands | 6.38 ± 0.94 $R_A$ |
| Abandoned Ridges | 6.08 ± 1.80 $R_A$ |

The average helium ratios, shown in Table 2, are only part of the overall picture; occluded helium measured in basalt from 'hotspots', such as Iceland and the Hawaiian Islands, typically have $^3$He/$^4$He ratios relative to air that are greater than 10 times the value in air with some samples measured as high as high as 37 $R_A$ [64].

Numerical simulations of georeactor operation, conducted at Oak Ridge National Laboratory, provide compelling evidence for georeactor existence: Georeactor helium fission products matched quite precisely the $^3$He/$^4$He ratios, relative to air, observed in oceanic basalt as shown in Figure 8. Note in that figure the progressive rise in $^3$He/$^4$He ratios over time as uranium fuel is consumed by nuclear fission and radioactive decay. The high $^3$He/$^4$He ratios observed in samples from 'hotspots' are consistent with the sharp increases observed from georeactor simulations as the uranium fuel becomes depleted and $^4$He diminishes.

Thermal structures, sometimes called mantle plumes, beneath the Hawaiian Islands and Iceland, two high $^3$He/$^4$He hot-spots, as imaged by seismic tomography [66, 67], extend to the interface of the core and lower mantle, further reinforcing their georeactor-heat origin. The high $^3$He/$^4$He ratios measured in 'hotspot' lavas appear to be the signature of 'recent' georeactor-produced heat and helium, where 'recent' may extend several hundred million years into the past. Recently, Mjelde and Faleide [68] discovered a periodicity and synchronicity through the Cenozoic in lava outpourings from Iceland and the Hawaiian Islands, 'hotspots' on opposite sides of the globe, that Mjelde et al. [69] suggest may arise from variable georeactor heat-production.

## 4.0 Georeactor Existence Evidence Based on Antineutrino Measurements

As early as 1930, it seemed that energy mysteriously disappeared during the process of radioactive beta decay. To preserve the idea that energy is neither created nor destroyed, 'invisible' particles were postulated to be the agents responsible for carrying energy away unseen. Finally, in 1956 these 'invisible' antineutrinos from the Hanford nuclear reactor were detected experimentally [70].

As early as the 1960s, there was discussion of antineutrinos being produced during the decay of radioactive elements in the Earth. In 1998, Raghavan et al. [71] were instrumental in demonstrating the feasibility of their detection. In 2002, Raghavan [72] authored a paper, entitled "Detecting a Nuclear Fission Reactor at the Center of the Earth" wherein he showed that antineutrinos resulting from nuclear fission products would have a different energy spectrum than those resulting from the natural radioactive decay of uranium and thorium. Raghavan's 2002 paper stimulated intense interest worldwide, especially with groups in Italy, Japan and Russia. Russian scientists [73] expressed well the importance: "Herndon's idea about georeactor located at the center of the Earth, if validated, will open a new era in planetary physics".

The georeactor is too small to be presently resolved from seismic data. Oceanic basalt helium data, however, provide strong evidence for the georeactor's existence [24, 62] and antineutrino measurements have not refuted its existence [74, 75]. To date, detectors at Kamioka, Japan and at Gran Sasso, Italy have detected antineutrinos coming from within the Earth. After years of data-taking, an upper limit on the georeactor nuclear fission contribution was determined to be either 26% (Kamioka, Japan) [75] or 15% (Gran Sasso, Italy) [74] of the total energy output of uranium and thorium, estimated from deep-Earth antineutrino measurements (Table 3). The actual total georeactor contribution may be somewhat greater, though, as some georeactor energy comes from natural decay as well as from nuclear fission.

**Table 3.** Antineutrino determinations of radiogenic heat production [74, 75] shown for comparison with Earth's heat loss to space [76]. See original report for discussion and error estimates.

| Heat (terawatts) | Source |
|---|---|
| 44.2 TW | global heat loss to space |
| 20.0 TW | antineutrino contribution from $^{238}$U, $^{232}$Th, and georeactor fission |
| 5.2 TW | georeactor KamLAND data |
| 3.0 TW | georeactor Borexino data |
| 4.0 TW | $^{40}$K theoretical |
| 20.2 TW | loss to space minus radiogenic |
| 15.0 TW | natural radioactivity of $^{238}$U, $^{232}$Th, $^{40}$K, without nuclear fission |

**5.0 Heat Flow Considerations and Georeactor Contributions to Geodynamics**

The 1940s concepts about Earth's interior are still evident in current assumption-based models: The inner core is still assumed to be partially crystallized iron metal, and the mantle is still assumed to be of uniform composition and considerably more oxidized than an enstatite chondrite. Mantle convection was added with the introduction of seafloor spreading and plate tectonics. Within that framework, computational models of "primitive mantle composition" continue to be promulgated. For example, the Bulk Silicate Earth (BSE) model attempts to model the unknown "primitive mantle composition" from carbonaceous chondrite element abundances, modified by inferences derived from the assumption that observed peridotite compositions result from partial melting of the assumed primitive mantle [77].

In BSE, uranium and thorium are assumed to reside entirely in the crust and mantle. Heat from the radioactive decay of $^{238}$U, $^{232}$Th and $^{40}$K is assumed to be delivered efficiently to the surface by mantle convection. But even if rapid delivery were possible, heat from natural radioactive decay without georeactor fission, ~15 TW, is insufficient to account for Earth's radiated energy [76] (Table 3), and mantle convection is problematic, both from the standpoint of heat transport and geodynamics [49, 78].

The Earth's mantle is bottom heavy, 62% denser at the bottom than at the top [46]. The small amount of thermal expansion at the bottom (<1%) cannot overcome the 62% higher density at the mantle's bottom; bottom mantle matter cannot float to the mantle top. Sometimes attempts are made to obviate the 'bottom heavy' prohibition by adopting the tacit assumption that the mantle behaves as an ideal gas, with no viscous losses, *i.e.*, 'adiabatic'. But the mantle is a solid that does not behave as an ideal gas as demonstrated by earthquakes occurring at depths as great as 660 km. Earthquakes in the upper mantle indicate the catastrophic release of stress, observations inconsistent with adiabatic absence of viscous loss.

Mantle silicates are thermal insulators. Absent mantle convection, heat flow to the surface may be extremely inefficient. Moreover, without mantle convection, plate tectonics lacks a valid scientific basis. But that should not be surprising as there are other problems. For example, nowhere in the literature of plate tectonics is presented a logical, causally related explanation for the fact that approximately about 41% of Earth's surface is continental rock (sial) with the balance being ocean floor basalt (sima). Absent mantle convection, there is no motive force for continuing sequences of continent collisions, thought to be the sole basis for fold-mountain formation. The reasonable conclusion therefore is that there must exist a new and fundamentally different geoscience paradigm which obviates the problems inherent in plate tectonics and in planetesimal Earth formation, and yet is capable of better explaining observed geological features.

I have disclosed a new indivisible geoscience paradigm, called Whole-Earth Decompression Dynamics (WEDD) [79], that begins with and is the consequence of our planet's early formation as a Jupiter-like gas giant (Figures 9 and 10) and which permits deduction of: (1) Earth's internal composition and highly-reduced oxidation state (Figure 5); (2) Core formation without whole-planet melting; (3) Powerful new internal energy sources, protoplanetary energy of compression and georeactor nuclear fission energy; (4) Mechanism for heat emplacement at the base of the crust [80]; (5) Nuclear fission georeactor geomagnetic field generation [26]; (6) Decompression-driven geodynamics that accounts for the myriad of observations attributed to plate tectonics without requiring physically-impossible mantle convection [30, 81]; and, (7) A mechanism for fold-mountain formation that does not necessarily require plate collision [82] (Figure 10).

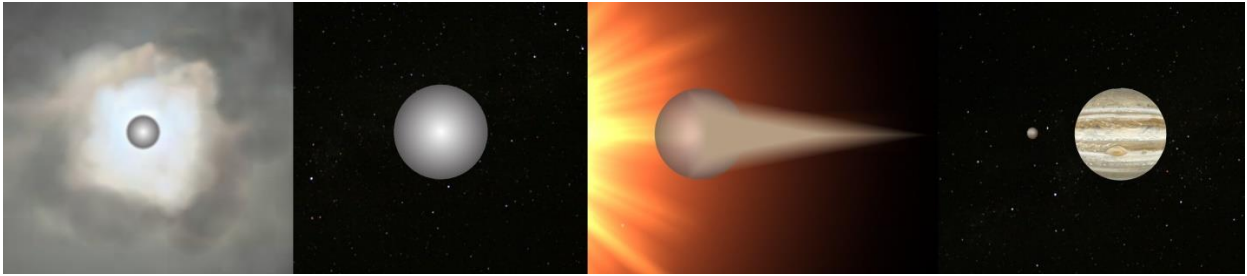

**Figure 9.** Whole-Earth Decompression Dynamics formation of Earth. From left to right, same scale: 1) Earth condensed at the center of its giant gaseous protoplanet; 2) Earth, a fully condensed a gas-giant; 3) Earth's primordial gases stripped away by the Sun's T-Tauri solar eruptions; 4) Earth at the onset of the Hadean eon, compressed to 66% of present diameter; 5) Jupiter for size comparison.

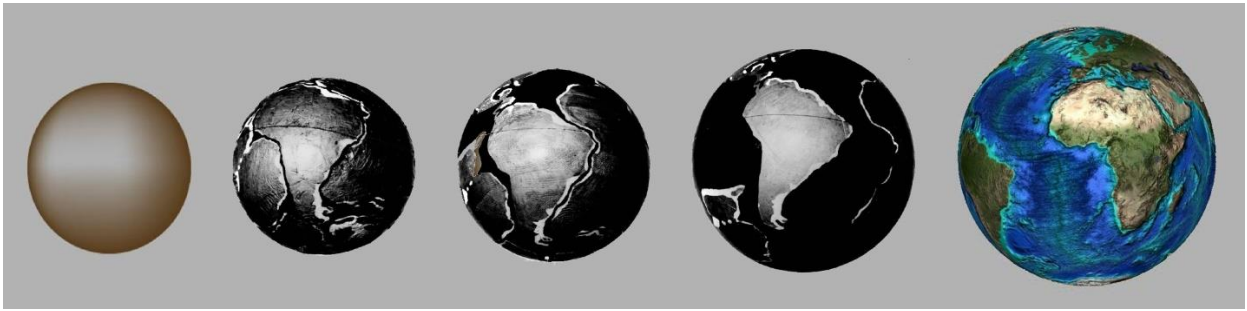

**Figure 10.** Schematic representation of the decompression of Earth (WEDD) from Hadean to present. From left to right, same scale: 1) Earth after T-Tauri removal of gases, 66% of present Earth diameter, fully covered with continental-rock crust; 2), 3), and 4) Formation of primary and secondary decompression cracks that progressively fractured the continental crust and opened ocean basins. Timescale not precisely established. 5) Holocene Earth. The geology of Earth, according to WEDD, is principally determined by Earth's decompression: Surface crack formation to accommodate increased planetary volume, and mountain formation to accommodate changes in surface curvature (Figure 11).

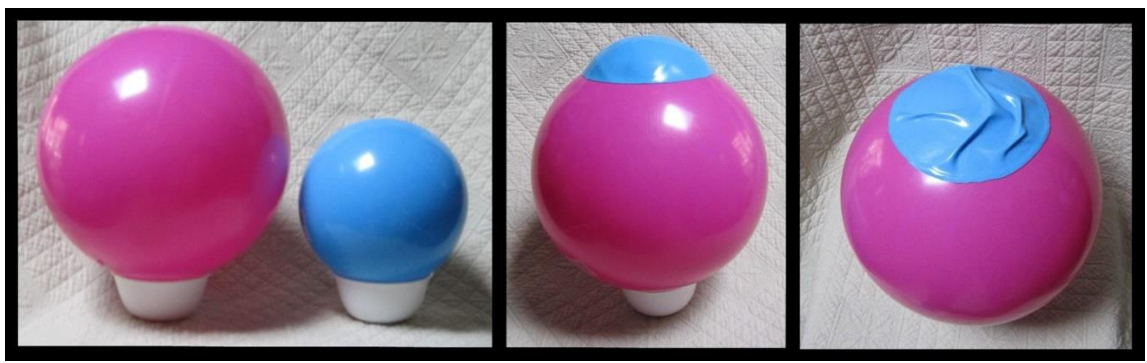

**Figure 11.** Demonstration illustrating the formation of fold-mountains as a consequence of Earth's early formation as a Jupiter-like gas giant. On the left, two balls representing the relative proportions of 'present' Earth (pink), and 'ancient' Earth (blue) before decompression. In the center, a spherical section, representing a continent, cut from 'ancient' Earth and placed on the 'present' Earth, showing: (1) the curvature of the 'ancient continent' does not match the curvature of the 'present' Earth and (2) the 'ancient continent' has 'extra' surface area confined within its fixed perimeter. On the right, tucks remove 'extra' surface area and illustrate the process of fold-mountain formation that is necessary for the 'ancient' continent to conform to the curvature of the 'present' Earth. Unlike the ball-material, rock is brittle so tucks in the Earth's crust would break and fall over upon themselves producing fold-mountains.

Heat from georeactor nuclear fission and radioactive decay can augment mantle decompression by replacing the lost heat of protoplanetary compression. The resulting decompression will tend to propagate throughout the mantle, like a tsunami, until it reaches the barrier posed by the base of the crust. There, crustal rigidity opposes continued decompression, pressure builds and compresses matter at the mantle-crust interface, resulting in compression heating, which, I posit, is the origin the heat responsible for the geothermal gradient [80].

Georeactor-produced heat, channeled to Earth's surface [49], provides a conduit through which highly mobile helium can readily migrate upward. Ultimately, the helium is occluded in volcanic lava produced by that heat. Seismically-observed heat channeling, beneath Iceland and the Hawaiian Islands, extends downward to the top of the core [66, 67]. The association of these 'hotspots' with helium, characterized by $^3He/^4He > 10\ R_A$, is consistent with the georeactor origin of that heat. Generalizing, high helium isotope ratios, $^3He/^4He > 10\ R_A$, may be taken as a signature of georeactor-heat in geological circumstances, even those in which the deep-extending seismic low-velocity profile is no longer evident, *e.g.*, the Siberian traps, which formed 250 million years ago [83].

Some mantle plumes, characterized occluded $^3$He/$^4$He > 10 $R_A$, appear to be associated with continent fragmentation [84]. Georeactor-produced heat is currently involved in the decompression-driven continent-splitting currently occurring along the East African Rift System and may be associated with the petroleum and natural gas deposits discovered there and in other similar circumstances [85]. Generalizing, georeactor-produced heat may augment decompression-driven rift formation that either splits continents or fails to split continents. For a more lengthy discussion, please see references [49, 50, 78, 85].

## 6.0 Georeactor Origin of the Geomagnetic Field

In 1939, Elsasser began a series of scientific publications in which he proposed that the geomagnetic field is produced within the Earth's fluid core by an electric generator mechanism, also called a dynamo mechanism [86-88]. Elsasser proposed that convection currents in the Earth's electrically-conducting iron-alloy core, twisted by Earth's rotation, act like a self-sustaining dynamo, a magnetic amplifier, producing the geomagnetic field. For decades, Elsasser's dynamo-in-the-core has been generally considered to be the only potentially viable means for producing the geomagnetic field and has been generally accepted without question. But, I discovered, there are serious problems, not with his idea of a convection-driven dynamo, but with its location within the Earth and with its energy source [26, 27, 49, 50].

There are periods in geological history when the geomagnetic field has been stable for millions of years. Such long-term convection stability cannot be expected in the Earth's fluid core. Not only is the bottom-heavy core about 23% denser at the bottom than the core-top, but the core is surrounded by a thermally insulating blanket, the mantle, with lower thermal conductivity and lower heat capacity than the core. Heat brought to the top of the core cannot be efficiently removed; the core-top cannot be maintained at a lower temperature than the core-bottom as necessary for thermal convection. Moreover, there is no obvious source of magnetic seed fields in the fluid core.

I have suggested that the geomagnetic field is produced by Elsasser's convection-driven dynamo operating within the georeactor's nuclear waste sub-shell [26]. Unlike the Earth's core, sustained convection appears to be quite feasible in the georeactor sub-shell (Figure 7). The top of the georeactor sub-shell is in contact with the inner core, a massive heat sink, which is in contact with the fluid core, another massive heat sink. Heat brought from the nuclear sub-core to the top of the georeactor sub-shell by convection is efficiently removed by these massive heat sinks thus maintaining the sub-shell adverse temperature gradient. Moreover, the sub-shell is not bottom-heavy. Unlike the fluid core, decay of neutron-rich nuclides in the nuclear waste sub-shell provides electrons that might form the seed magnetic fields for amplification.

## 7.0 Georeactor Geomagnetic Reversals

From time to time, on an irregular basis, the geomagnetic field reverses; North becomes South and vice versa. The average time between reversals is about 250,000 years; the last reversal of the geomagnetic field occurred about 750,000 years ago. There are, however, periods of time up to 50 million years in length when the Earth's magnetic field did not reverse at all.

Reversals of the geomagnetic field are produced when stable convection is interrupted in the region where convection-driven dynamo action occurs, in the nuclear waste sub-shell of the georeactor. Upon re-establishing stable convection, convection-driven dynamo action resumes with the geomagnetic field either in the same or in the reversed direction. The mass of the georeactor is quite low, less than one ten-millionth the mass of fluid core. Consequently, reversals can occur much more quickly, and with greater ease, than previously thought.

Trauma to the Earth, such as a massive asteroid impact or the violent splitting apart of continental land masses, might de-stabilize georeactor dynamo-convection, causing a magnetic reversal or excursion. It is also possible that a geomagnetic reversal might be caused by a particularly violent event on the Sun.

Earth is constantly bombarded by the solar wind, a fully ionized and electrically conducting plasma, heated to about 1 million degrees Celsius, that streams outward from the Sun and assaults the Earth at a speed of about 1.6 million kilometers per hour. The geomagnetic field deflects the brunt of the solar wind safely past the Earth, but some charged particles are trapped in donut-shaped belts around the Earth, called the Van Allen Belts. The charged particles within the Van Allen Belts form a powerful ring current that produces a magnetic field that opposes the geomagnetic field near the equator. If the solar wind is constant, then the ring current is constant and no electric currents are transferred through the magnetic field into the georeactor by Faraday's induction. High-intensity changing outbursts of solar wind, on the other hand, will induce electric currents into the georeactor, causing ohmic heating in the sub-shell, which in extreme cases might disrupt convection-driven dynamo action and lead to a magnetic reversal or excursion.

From ancient lava flows, scientists have recently confirmed evidence of episodes of rapid geomagnetic field change – six degrees per day during one reversal and another of one degree per week – were reported [89, 90]. The relatively small mass of the georeactor is consistent with the possibility of a magnetic reversal occurring on a time scale as short as one month or several years.

Nuclear fission consumes uranium at a much faster rate than natural radioactive decay. At some unknown time in the future, disruption of convection-driven dynamo action will occur, but unlike a magnetic reversal or excursion, there will be insufficient uranium remaining to re-establish stable convection and stable geomagnetic field production [24].

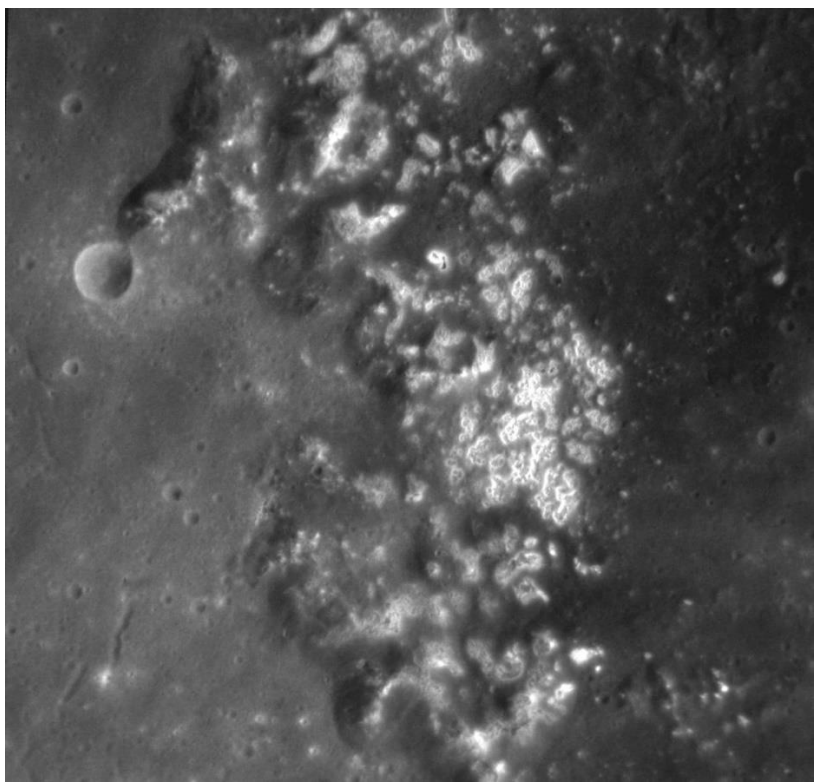

**Figure 12.** NASA MESSENGER image, taken with the Narrow Angle Camera, shows an area of hollows on the floor of Raditladi basin on Mercury. Surface hollows were first discovered on Mercury during MESSENGER's orbital mission and have not been seen on the Moon or on any other rocky planetary body. This high-reflectance material, I posit, formed as hydrogen from Mercury's core exited the planet and reduced iron sulfide to the metal. Verifying that the bright metal is in fact low-nickel iron metal will provide strong evidence for high-pressure, high-temperature protoplanetary formation, and concomitant planetocentric nuclear reactor formation.

## 8.0 Perspectives

The chemical state and location of thorium in the deep-interior of Earth is unknown. Like uranium, thorium occurs mainly in that portion of the Abee enstatite chondrite that corresponds to the Earth's core [45]. But very little thorium can exist within the nuclear sub-core; otherwise, the absorption of neutrons by thorium would poison the nuclear fission chain reaction. So, a fundamental question arises as to whether thorium resides in the georeactor nuclear waste sub-shell or elsewhere in the core. If thorium resides in the georeactor nuclear waste sub-shell, then the question becomes whether thorium interaction with neutrons will breed additional fuel, *i.e.*, fissionable $^{233}$U.

Detection of antineutrinos, also called geoneutrinos, which originate deep within the Earth has the potential for verifying georeactor existence and for revealing the location of uranium and thorium. Already geoneutrino detection has established upper limits on the georeactor nuclear fission contribution of either 26% [75] or 15% [74] of the total energy output of uranium and thorium. Improving detection efficiency and lowering background will be beneficial, but will not delineate actinide locations; directional detectors will be required, as well as verification that geoneutrino attenuation with depth is insignificant as presently believed.

The georeactor is not unique to Earth. Its existence is the consequence of planetary formation by raining-out at high temperatures and pressures in the interior of a giant gaseous protoplanet. The commonality of that mode of planetary formation, yielding highly reduced condensed in the Solar System, and the commonality of micro-gravity georeactor operating environments makes understandable georeactor-type magnetic field generation in most planets and large moons [27, 50].

Many of the images from the MESSENGER spacecraft, like Figure 12, reveal "… an unusual landform on Mercury, characterized by irregular shaped, shallow, rimless depressions, commonly in clusters and in association with high-reflectance material … and suggest that it indicates recent volatile-related activity" [91]. But the authors, reasoning within the framework of consensus-approved models, were unable to describe a scientific basis for the source of those volatiles or to suggest identification of the "high-reflectance material". By contract, I [92] calculated that, during condensation at pressures ≥1 atm., copious amounts of hydrogen, one or more Mercury volumes at STP could be incorporated in Mercury's fluid iron alloy core, which will be released as the core subsequently solidifies. Hydrogen geysers, exiting the surface, I posited, formed the pits or hollows and are possibly involved in the exhalation of iron sulfide, which is abundant on the planet's surface, and some of which may have been reduced to iron metal thus accounting for the associated "high-reflectance material", bright spots. Verifying that the "high-reflectance material" is indeed metallic iron of low nickel content will not only provide strong evidence for Mercury's hydrogen geysers, but more generally will provide evidence that planetary interiors "rained out" by condensing within giant-gaseous protoplanets at high pressures, which is the same circumstance responsible for causing uranium to reside in the core and function as a planetocentric nuclear fission reactor.

The agreement between observed $^3$He/$^4$He in deep-source oceanic basalt and the numerical simulation helium fission products is strong evidence for georeactor existence (Section 3.0). The calculated progressive increase in the ratio over time (Figure 8) is particularly significant in light of $^3$He/$^4$He > 10 $R_A$ observed in hotspot "mantle plumes" that have been seismically imaged to the edge of the core. It should be possible in principle to observe the change in $^3$He/$^4$He over time by measurement along volcanic chains, such as the Hawaiian-Emperor volcanic chain, as was initially measured by Keller et al. [93]. Their results suggest a progressive increase in $^3$He/$^4$He since the Cretaceous; clearly further work is needed for better statistics and to more clearly delineate mineral hosts that best preserve the helium signature over time.

Nuclear fission consumes uranium fuel; eventually, the georeactor fuel supply will become depleted, thermal convection will cease in nuclear waste sub-shell, and the geomagnetic field will collapse [24]. When will georeactor demise occur? That is currently unknown, but can in principle and should be addressed through the measuring the temporal change in deep-source helium ratios. It is important to determine the time of georeactor demise, because of the low mass of the georeactor, one ten-millionth that of the core, the end may come quickly with potentially devastating effects on humanity.

## Acknowledgements

I have benefitted from conversations with P. K. Iyengar, Paul K. Kuroda, Inge Lehmann, Lynn Margulis, Hans E. Suess, and Harold C. Urey. I thank Daniel F. Hollenbach and Oak Ridge National Laboratory personnel who generously devised a methodology to allow their software, originally designed for man-made nuclear reactors, to utilize geological time scales and to simulate fission product removal [28].